\documentclass[12pt]{article}
\usepackage{epsfig}
\usepackage{amsmath}
\begin{document}

 \title{ On the Question of Interference in Radiation Produced by Relativistic Channeled Particles}
 \author{\sl V.F.Boldyshev and M.G.Shatnev
 \footnote{E-mail:shatnev@kipt.kharkov.ua} ,
 \\ National Science Center\\ Kharkov Institute of
Physics and Technology,\\
 61108, Kharkov, Ukraine.}
 \date{}
 \maketitle
 \begin{abstract}
 Two approaches used in the description of the channeling radiation
 emitted from relativistic positrons are compared with each other.
 In the first (traditional) case, the probability of the process is proportional
 to a sum of absolute squares of the amplitudes of the transition
 between two states with definite transverse energy levels of the positrons
 traversing single crystals. In the second case, we begin with calculation of the sum of
 amplitudes for transition between states  with different transverse energy levels
 for corresponding radiation frequency, and then the sum is squared. One must keep in mind
 that the latter approach can be used only in the case when positrons  move
  in a nearly harmonic planar potential with equidistant transverse energy levels.
  It is shown that the calculation based on the second approach can give rise
   to a peak structure in the spectrum when the number of transverse energy levels
   is much greater than one.
 \end{abstract}

      We consider a relativistic charged particle incident onto a crystal at a small angle
       to a crystal planes. In the planar channeling case for positively charged positrons,
       the channel is between the crystal planes. This channel is the source of
        a potential well in the direction transverse to particle motion giving rise
        to transversely bound states for the particle. Transitions
         to lower-energy states lead to the phenomenon known as channeling radiation (CR).
     Calculations of the CR process are carried out by using the rules of quantum
      electrodynamics[1,2]. The longitudinal and transverse motion of relativistic
       channeled positrons is described by the time-independent Dirac equation and
        the one-dimensional Schr\"{o}dinger  equation  with relativistic corrections,
        respectively. The wave functions in the matrix element are given by the solutions
        of these equations, and, since the potential in our case only depends on the coordinate x,
         the part of the wave function describing the free motion along the y- and z-axes is
          a plane wave. The part of the wave function describing transverse nonrelativistic
        one-dimensional oscillations of planarly channeled positrons in the laboratory frame
        is obtained solving the Schr\"{o}dinger  equation for the transverse motion.
       The intensity of the CR is proportional to the square of the matrix element absolute value.
 The Doppler formula for the energy of photon emitted is derived using the energy
 and momentum conservation laws. For this energy one gets

\begin{equation}\label{eq1}
  \omega =\frac{\varepsilon _n-\varepsilon _{n^{\acute{}}}}{1-\frac{p_{\Vert }%
}E\cos \theta } ,
 \end{equation}
where $\varepsilon _n$ and $\varepsilon _{n^{\acute{}}}$ are the
discrete energy levels of the transverse oscillations of the
positron in the channel before and after radiation, respectively,
$E$ and $p_{\Vert }$ are the energy and the longitudinal momentum
of the positron, and $\theta $ is the angle of radiation emission
relative to the direction of motion of the channeled positron.

The radiation of a maximum frequency
$$\omega =2(\varepsilon _n-\varepsilon _{n^{\acute{}}})\gamma ^2,   \gamma =\frac Em$$
is emitted in the forward direction (at $\theta =0$ ). For
positrons of not too high transverse energies, a good
approximation (see e.g., Refs.[1-3]) is the harmonic potential
leading to equidistant energy levels

$$\varepsilon _n=\Omega (n+1/2),$$
where $\Omega =\frac 2{d_p}\sqrt{\frac{2U_0}E}$ is the oscillation
frequency, $d_p$ is the distance between planes in the
corresponding units, and $U_0$ is the depth of the potential well.
Since $%
\frac{p_{\Vert }}E=(1-\frac 12\gamma ^{-2}),$ and $\cos \theta
=(1-\frac 12\theta ^2),$  Eq. (2) can be expressed as
\begin{equation}\label{eq2}
  \omega =2\gamma ^2(n-n^{\acute{}})\Omega (1+\theta ^2\gamma
  ^2)^{-1}.
 \end{equation}

The case $n-n^{\acute{}}=1$ corresponds to the peak values of the
experimental CR spectra [4], being the first harmonic with the
photon energy $\omega =2\gamma ^2\Omega $. As it follows from
Eq.(2), photons emitted via positron transition from any initial
level $n$ to the final level $  n-1$ are identical (i.e., have the
same energies for the same emission angles). This means that the
resulting amplitude should be given by an additive superposition
of amplitudes of all such transitions. The positron state outside
the crystal  $(z<0)$ is a plane wave, whereas inside the crystal
$(z>0)$ , the part of its wave function corresponding to the
transverse motion is a superposition of the harmonic oscillator
eigenvectors. Factors $c_n$ describing transitions from the
initial state to states with the transverse energy levels $n$ can
be found using boundary conditions set upon the wave function at
the crystal border $(z=0)$. Then, a transition to the closest
lower level $n-1$ occurs with emission of a photon having energy
$\omega $. The amplitude $M_{n,n-1}$ of such a transition was
calculated in Refs.[1,2]. One may expect that the total amplitude
 of the transition from the initial to final state accompanied by
 the photon emission is determined by products of the amplitudes
  $c_n$ and $M_{n,n-1}$. Following the rules of the quantum
  mechanics, we express this amplitude as

  \begin{equation}\label{eq3}
\\A \propto\sum_{n}c_nM_{n,n-1} ,
\end{equation}
were summation is performed over all the harmonic oscillator
levels. In paper [1] the spectral-angular distribution of emitted
photons is represented as

\begin{equation}\label{eq4}
\\\frac{d^2\varpi }{d\omega dO} \propto\sum_{f}\left| M_{if}\right|
^2.
\end{equation}

The sum entering Eq.(4) is the one over the quantum numbers f of
the transverse motion of the particle. Then, the probability of
having a definite transverse energy is taken into account by
multiplying each term of this sum by a corresponding factor. In
our consideration, discrete levels of the transverse motion in the
harmonic oscillator potential refer to the intermediate state of
the particle. Accordingly, the contribution to the intensity
$\frac{d^2\varpi }{d\omega dO}$ due to transitions, e.g., between
the closest levels is determined by the square of the absolute
value of expression (3)

\begin{equation}\label{eq5}
\\\frac{d^2\varpi^{(1)} }{d\omega dO}\propto\left|\sum_{n}c_nM_{n,n-1}\right|
^2.
\end{equation}

In other words, unlike Ref.[1], we get an expression that contains
interference terms mixing amplitudes of photon emission from
different equidistant levels. We would like to note that dynamics
of channeling electron in a crystal differs from that of the
positron case. The transverse potential well for the electron does
not give rise to equidistant energy levels for transverse particle
motion. Therefore, there are no interference contributions to the
photon emission intensity similar to those present in Eq. (5). In
our opinion, this could explain the greater intensity in case of
channeling positron compared to that for the electron observed in
experiment [4]. We think that studying the problems like the one
under consideration in this paper can help in the development of
methods for obtaining polarized photon beams.

The authors would like to thank L.G. Levchuk for helpful
discussions and interest to the work.

 \end{document}